# Method for measuring prompt γ-rays generated by D-T neutrons bombarding a depleted uranium spherical shell[*]

QIN Jian-Guo(秦建国)[1,2,3]  LAI Cai-Feng(赖财锋)[2]  JIANG Li(蒋励)[2]  LIU Rong(刘荣)[2;1)]
ZHANG Xin-Wei(张信威)[4]  YE Bang-Jiao(叶邦角)[1]  ZHU Tong-Hua (朱通华)[2]

[1] Department of Modern Physics, University of Science and Technology of China, Hefei 230026, China
[2] Institute of Nuclear Physics and Chemistry, CAEP, P.O. Box 213, Mianyang 621900, China
[3] Graduate School, China Academy of Engineering Physics, Mianyang 621900, China
[4] Institute of Applied Physics and Computational Mathematics, P.O. Box 8009, Beijing 100088, China

**Abstract:** The prompt γ-ray spectrum from depleted uranium (DU) spherical shells induced by 14 MeV D-T neutrons is measured. Monte Carlo (MC) simulation gives the largest prompt γ flux with the optimal thickness of the DU spherical shells 3−5 cm and the optimal frequency of neutron pulse 1 MHz. The method of time of flight and pulse shape coincidence with energy (DC-TOF) is proposed, and the subtraction of the background γ-rays discussed in detail. The electron recoil spectrum and time spectrum of the prompt γ-rays are obtained based on a 2″ × 2″ BC501A liquid scintillator detector. The energy spectrum and time spectrum of prompt γ-rays are obtained based on an iterative unfolding method that can remove the influence of γ-rays response matrix and pulsed neutron shape. The measured time spectrum and the calculated results are roughly consistent with each other. Experimental prompt γ-ray spectrum in the 0.4−3 MeV energy region agree well with MC simulation based on the ENDF/BVI.5 library, and the discrepancies for the integral quantities of γ-rays of energy 0.4−1 MeV and 1−3 MeV are 9.2% and 1.1%, respectively.

## 1 Introduction

Depleted uranium is widely used in fields such as weapons physics, fusion-fission hybrid reactors and nuclear physics [1−3]. When a nuclear explosion occurs, the prompt radiation measurement [4−7] is an essential diagnostic method. In the design of hybrid-reactor physics, the radiation effects and $^{238}$U nuclear parameters must be taken into account. In addition, prompt γ-ray spectra [8, 9] are widely employed in the study of nuclear structure. Accurate cross sections for $^{238}$U and $^{235}$U with neutrons are important data in the design of nuclear facilities and the application of nuclear technology. Therefore, to investigate the prompt γ-rays emitted by D-T neutron bombard measurement with DU spherical shells has significant scientific meaning and potential applications.

As far as $^{235}$U and $^{238}$U are concerned, the fission cross sections are very accurate, but the radiation capture cross sections and inelastic cross sections have to be refined. In different nuclear databases, there are prominent discrepancies in certain energy regions. Significant deviations and inadequacies exist in the experimental data in the high-energy region. For the inelastic scattering cross sections of $^{238}$U induced by 14 MeV neutrons, the differences between CENDL3.1, JEFF3.1.2 and JENDL4.0 [10, 11 ,12] databases are about 12%, 9% and 6% relative to ENDF/B7.1 [13], respectively, and the discrepancies between the experimental value and the evaluated nuclear databases are more serious [14]. Measurement of the prompt γ-ray spectrum is one important method to validate the cross sections [15−17].

[*] Supported by the National Special Magnetic Confinement Fusion Energy Research, China, under contract No. 2015GB108001 and National Natural Science Foundation of China (No．91226104).
1) *E-mail*: stingg@126.com

The benchmark data obtained in integral experiments with macroscopic samples can be used to evaluate cross sections and check databases. One such experiment is the pulsed spheres experiment [18−21] conducted in LLNL in 1989, in which the electron recoil spectra (ERS) of $^{238}$U spherical shell sample were investigated under D-T neutron irradiation. The inner and outer radii of the uranium sphere shell were 3.63 cm and 10.91 cm, respectively, and the energy of the γ-rays was between 0.5 and 8 MeV. The experimental results were higher by about 6% compared with computations. In 1990, the ERS of 0.17 to 6 MeVee was measured, and the ERS agrees well with the simulation below 4 MeVee [19], but the details of the γ-ray time spectrum were not reported. Other highly−cited experiments were performed on the Si, W, Pb, Cu, V and other benchmark assemblies at the Japan Atomic Energy Institute and Osaka University from 1989 to 1994 [22−24]. The prompt γ-rays from the assemblies bombarded by D-T neutron were studied with liquid scintillation detectors, while fissionable material was not included. In China, leakage γ-ray spectra from boron polyethylene and iron sphere have been measured at the China Academy of Engineering Physics (CAEP) [25, 26]. However, only the γ-ray energy spectrum in the $^{238}$U spherical shell was measured [27], and the prompt γ-ray energy spectrum and time spectrum were not presented.

In this paper, the method of measuring prompt γ-rays from a DU spherical shell with pulsed deuterium-tritium reaction neutrons is introduced, based on the technique called "double-coincidence and time of flight" (DC-TOF), and the preliminary experimental results are compared with Monte Carlo simulation.

## 2 Characteristics of prompt γ-rays from DU

Inelastic scattering, radiation capture and fission reaction in some nuclear materials under neutron irradiation will emit prompt γ-rays (namely inelastic scattering γ-rays), radiation capture γ-rays and prompt fission γ-rays, respectively. The emission of prompt γ-rays is as fast as $10^{-14}$ to $10^{-7}$ seconds [28−30]. The daughter products of an excited or metastable nucleus will de-excite to the ground state and emit γ-rays, the so-called delayed γ-rays. The decay half-life of fission nuclide is usually longer than one second, while the shortest is only 0.032 seconds [28]. Therefore, prompt γ-rays and decay γ-rays or delayed γ-rays can be distinguished via the emission time.

The sample here is a DU spherical shell with inner and outer diameters of 50.8 cm and 57 cm, respectively. The DU is composed of $^{238}$U, $^{235}$U, $^{234}$U and $^{236}$U. The prompt γ-rays from the sample bombarded by 14 MeV neutrons are mainly inelastic scattering γ-rays, prompt fission γ-rays and negligible radiation capture γ-rays. The characteristics of prompt γ-rays from $^{238}$U and $^{235}$U nuclei under neutron irradiation are shown in Table 1. The average energy <E> and multiplicity <G> of fission γ-rays were calculated by Eqs. (1) and (2) [31], <E> is in MeV.

$$<E> = -1.33(\pm 0.05) + 119.6(\pm 2.5) Z^{\frac{1}{3}} / A \tag{1}$$

$$<G> = \frac{E_t(\bar{v}, Z, A)}{<E>} \tag{2}$$

Table 1. Characteristics of prompt γ-rays produced by neutrons bombarding a DU spherical shell.

| Nuclide | Process | Energy region | Intensity distribution | Comments |
|---|---|---|---|---|
| $^{238}$U | fission | 0.25–7 MeV | 0.25–4.25MeV: > 95% <br> A quasi Maxwell distribution | $\sigma_{fission-14\,MeV}$ = 1.1 barn <br> $<G>$ = 7.46 <br> $<E>$ = 0.94 MeV |
| | inelastic | 44.91 keV–5.2 MeV | A quasi Maxwell distribution | $\sigma_{inl-14\,MeV}$ = 0.58 barn <br> 88 energy levels <br> Hundreds of γ-rays |
| | capture | 11.98 keV–4.806 MeV | 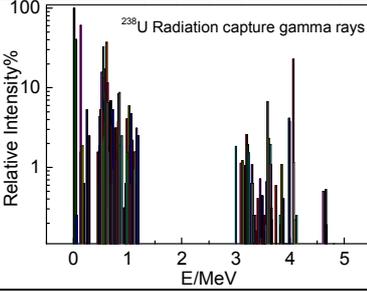 | $\sigma_{c-14\,MeV}$ = 9.8 × 10$^{-4}$ barn <br> < 0.4 MeV: 10 γ-rays <br> 0.4–1 MeV: 64 γ-rays <br> > 1 MeV: 94 γ-rays |
| $^{235}$U | fission | 0.25–7 MeV | < 2 MeV: 90.3% <br> A quasi Maxwell distribution | $\sigma_{fission-14\,MeV}$ = 2.0 barn <br> $<G>$ = 6.73 <br> $<E>$ = 0.97 MeV |
| | inelastic | 13 keV–1.14 MeV | A quasi Maxwell distribution | $\sigma_{inl-14\,MeV}$ = 0.4 barn <br> More than one hundred energy levels. <br> Hundreds of γ-rays and the most of energy less than 1 MeV. |
| | capture | 42.54 keV–6.5 MeV | 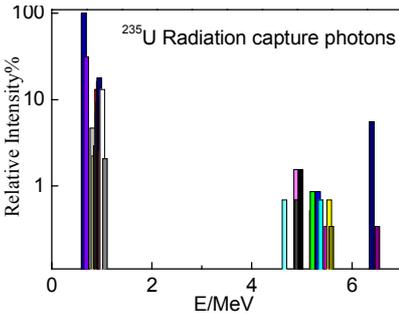 | $\sigma_{c-14\,MeV}$ = 1.2 × 10$^{-3}$ barn <br> 58 γ-rays. <br> The strongest γ-ray with energy of 642.2 keV. |

For the macroscopic sample here, the fission prompt γ-rays cannot be identified by fission fragments signal coincidence measurement as for a microscopic sample. Therefore, the prompt γ-rays discussed in this paper include fission prompt γ-rays and inelastic γ-rays, and the negligible radiation capture γ-rays.

## 3 Experiment
### 3.1 Experimental setup

The prompt γ-ray spectrum was measured using the "Dual Coincidence-TOF" method (the coincidence of γ pulse with flight time and the energy spectrum measured by a liquid scintillation detector) based on pulsed neutrons. The parameters of the uranium shell are shown in Table 2. The schematic of the experimental set-up is shown in Fig. 1. The distance of the BC501A detector to the tritium target is 10.7 m and the thickness of the concrete wall is 2 m. The shielding collimation hole is composed of iron and polyethylene with thicknesses of 5 cm and 40 cm, respectively.

Table 2. Description of the DU spherical assemblies.

| Material | $\rho$(g/cm$^3$) | $\rho\Delta R$*/(g/cm$^2$) | $R_{inner}$/cm | $R_{outer}$/cm | MFP** | Weight/kg |
|---|---|---|---|---|---|---|
| Depleted uranium*** | 18.8 | 51.6 | 25.4 | 28.5 | 0.87 | 526.6 |

\* Areal density of the assembly material.
\*\* Number of mean-free-paths calculated for 14 MeV neutrons.
\*\*\* Composition: $^{238}$U-99.579%, $^{235}$U-0.4154%, $^{234}$U-0.034%, $^{236}$U-0.003%.

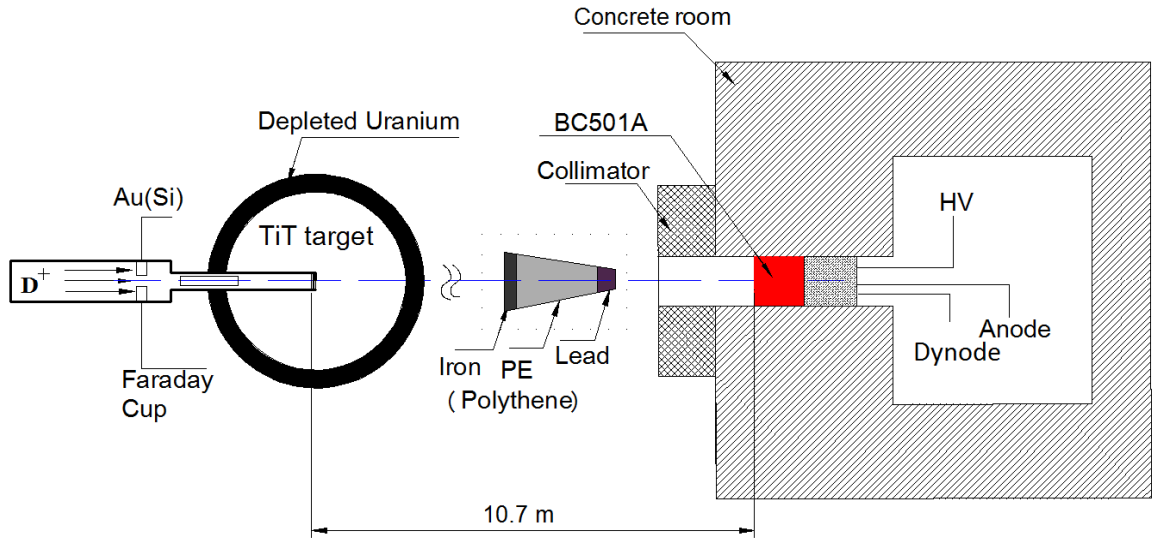

Fig. 1. Schematic of experimental setup based on liquid scintillation detector using TOF method.

## 3.2 Experiment parameters

In this work, the prompt γ-rays are mainly influenced by the delayed γ-rays from the activation nucleus and the fission products. For a sample irradiated by neutrons in square waveform mode shown in Fig. 2(a), the delayed γ-ray intensity will increase with the irradiation time and reach an equilibrium value after about five half-life irradiations. For the neutron source in pulse mode shown in Fig. 2(b), the delayed γ-rays can be subtracted very well. After neutron irradiation in pulse mode for a certain time, the delayed γ-ray will reach an equilibrium state. In one time period, prompt γ-rays will produced when the neutron pulse generated, and then only the delayed γ-rays are left after the neutron pulse. Therefore, we can obtain the prompt γ-rays by subtracting the delayed γ-rays from the total signal in one time period.

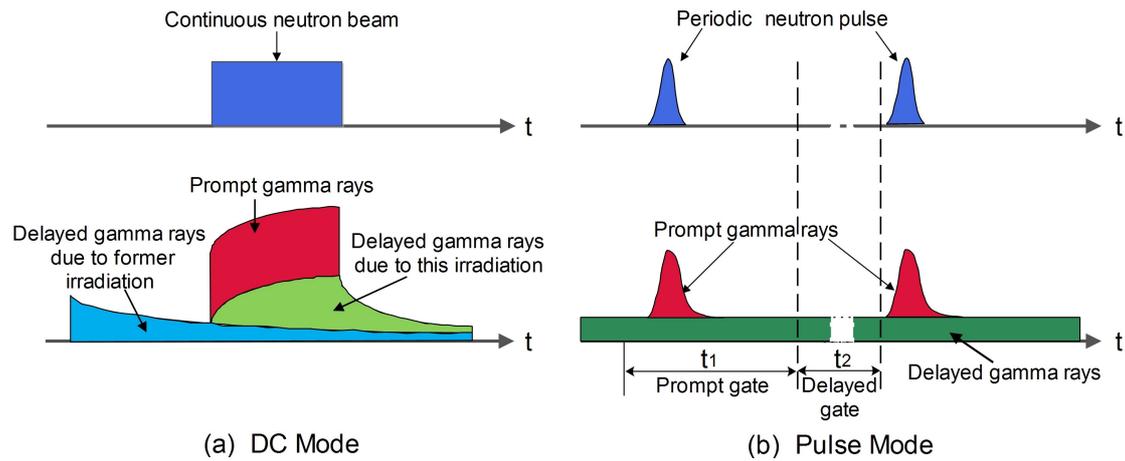

Fig. 2. Subtraction of delayed γ-rays produced by different waveforms of neutron source: (a) square waveform mode, and (b) pulse mode.

With the MCNP5 MC simulation, the optimal thickness of DU spherical shell and the accelerator parameters were determined. In the calculation, neutrons are emitted from a 14.9 MeV isotropic point source, located in the center of the spherical shell. The cross sections of neutron interaction with $^{14}$N and $^{16}$O are cited from the ENDF/VI.8 (actia); data for $^{238}$U and $^{235}$U are cited from the ENDF/BVI.5 Library (endf66c) and data for $^{234}$U and $^{236}$U are cited from the ENDF/BVI.0 Library (endf66c). Cross sections for photon and electron produced from C, O and U are cited from ENDF/BVI.8 Library (mcplib04 and el03). The total composition of $^{234}$U and $^{236}$U is less than 0.05% and their influence can be neglected.

The prompt γ-ray intensity, energy spectrum and time spectrum computed are shown in Fig. 3. The prompt γ-ray flux (see Fig. 3(a)) reaches the maximum value when the thickness of DU increases to 3 cm, and it changes very little with thickness up to 5 cm. We choose a spherical shell with inner and outer diameters of 50.8 cm and 57 cm, respectively, from available samples. In the energy spectra with different intervals shown in Fig. 3(b), a characteristic peak with energy 4.06 MeV emerges after 100 ns, which indicates that the $^{238}$U(n,γ)$^{239}$U reaction occurred. Most of the prompt γ-ray emission finishes within 100 ns as illustrated in Fig. 3(c). The radiation capture γ-rays, which are induced by low energy neutrons, extend beyond 100 ns. The time spectrum in Fig. 3(c) shows that γ-ray flux gradually increases in the first 4 ns interval, which is generated by neutrons bombarding the air in the shell. Therefore, 4 ns correspond to neutron arrival time from the inner surface to the outer surface of the spherical shell. The γ-ray flux decreases by 3 orders of magnitude after 200 ns, and it decreases by approximately 6 orders of magnitude after 1μs.

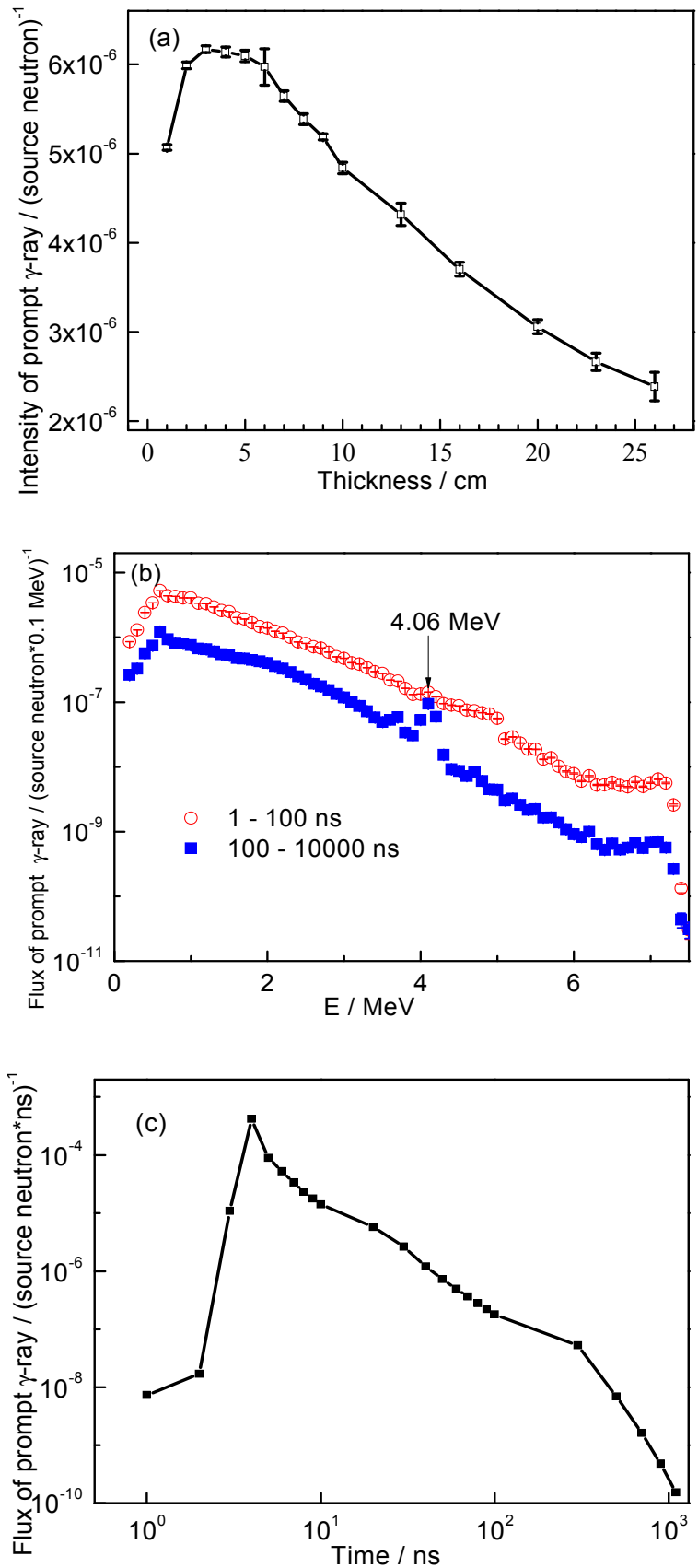

Fig. 3. Simulation results of prompt γ-rays produced by the DU spherical shell irradiated by pulsed neutrons using program MCNP5: (a) the γ-ray flux versus sample thickness, (b) energy spectra with different time intervals, and

(c) time spectrum of the leakage γ-rays.

Based on the simulation, it is believed that the prompt γ leakage rate is less than 0.1% when the intensity decreases by 3 orders of magnitude, in considering of the experimental detection efficiency. Therefore, the effective time of neutron interaction with the sample is less than 200 ns, which is also the emission period of prompt γ-rays. In order to distinguish the prompt from delayed γ-rays, the delayed γ-rays must be measured as background within less than 200 ns. Considering the parameters of the PD-300 neutron generator at the Institute of Nuclear Physics and Chemistry (INPC), Mianyang, China, neutrons in pulse mode with 10 ns of width and 1 MHz of frequency were used in the experiments described below.

### 3.3 Background subtraction

In this work, γ-rays were always generated with neutrons or even no neutron. Many factors may interfere with the prompt γ-ray measurement, most of which are described here: (1) delayed γ-rays emitted from the DU shell; (2) prompt γ-rays from the tritium target, material of target chamber and air in the shell under neutron irradiation; (3) γ-rays from the detector and its surrounding material bombarded by scattered neutrons; (4) prompt γ-rays from the detector generated by penetrating neutrons; (5) prompt γ-rays from assemblies in which the DU was excluded; (6) spontaneous fission γ-rays from the sample; (7) intrinsic γ-rays of the detector. The corresponding methods of subtracting these backgrounds are listed in Table 3.

Table 3. Methods of subtracting various γ-ray backgrounds.

| Background items | Method | Comments |
|---|---|---|
| 1. Delayed γ-rays from the DU sample | Pulsed neutron source | Key factors |
| 2. Prompt γ-rays from the detector generated by penetrating neutrons | Time of Flight | |
| 3. Prompt γ-rays from the tritium target, material of target chamber and air in the shell under neutron irradiation | Experiment & Simulation | Insignificant factors |
| 4. The γ-rays from the detector and its surrounding material bombarded by scattered neutrons | Time of Flight | |
| 5. Prompt γ-rays from experimental setup exclude the DU sample | Shadow cone | |
| 6. Spontaneous fission γ-rays | Pulsed neutron source | |
| 7. Intrinsic γ-rays of the detector | | |

The basic principle to subtract the delayed γ-rays is described in Section 2; here we give a brief summary on subtracting signals from other sources. Prompt γ-rays from the target and the uranium shell are detected almost at the same time (less than 2 ns), so we cannot discriminate these γ-rays either by emission time or by energy. However, we can measure the γ-rays of the target without the uranium shell, and then conduct simulation to determine the probability of the γ-rays penetrating the shell and the γ signals recorded by the detector. This method is listed in Table 3 as 'Time of Flight'. For the γ-rays from that experimental setup materials exclude the DU sample, a shielding cone (see Fig. 1) was placed between the uranium shell and the detector. This cone could shield the γ-rays from the shell, and let only the prompt γ-rays from the surrounding materials be recorded by the detector, which could then be subtracted in data processing.

The time spectrum of γ-rays based on the TOF method is shown in Fig. 4, AB, CD and EF are three time intervals marked as 1, 2 and 3, respectively. The γ-rays in period 1 were emitted from the uranium shell and the target, and are the first to reach the detector. The γ-rays in period 2 were generated from the detector and surrounding materials bombarded by the leakage neutrons. The time shift between the two peaks in periods 1 and 2 represents the flight time difference of these two types of γ-rays. Therefore, the prompt γ-rays in period 1 and the delayed γ-rays in period 3 can be measured, and the contribution to period 1 from delayed γ-rays can be deduced with a normalization time factor.

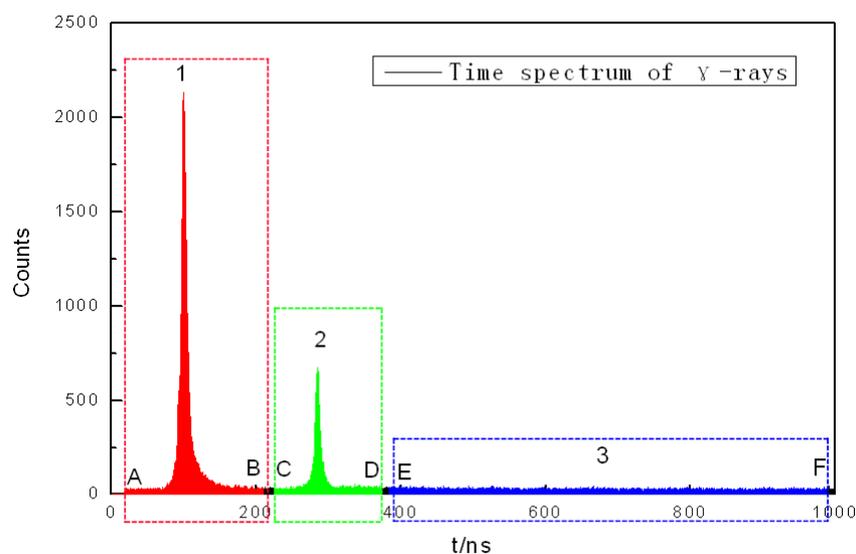

Fig. 4. Schematic of subtracting γ-ray background based on TOF method.

## 3.4 Neutron source and detectors

Referring to Fig. 1, 14 MeV neutrons were generated in the $^3$H(d,n)$^4$He reaction using the 205 keV D$^+$ beam from the PD-300 Cockcroft-Walton accelerator at INPC. The tritium target with 0.8 mm-thick molybdenum substrate was placed at the center of the uranium shell, in which the tritium is absorbed in a titanium layer with diameter 1.2 cm and mass thickness 1.34 mg/cm$^2$. The D$^+$ beam was swept and bunched with a repetition rate of 1 MHz and a burst width of ~ 10 ns. A solid-state Au(Si) surface barrier detector was placed at 174° with respect to the D$^+$ beam line, to monitor the neutron production by counting the associated α particles. The Faraday cup was located in the drift tube and so that the deuterium ions would pass through it and reach the tritium target. The 2″ × 2″ BC501A liquid scintillator detector was placed in a concrete room at 0° with respect to the D$^+$ beam line.

The maximum energy of the prompt γ-rays emitted from the DU shell is less than 7 MeV. The energy spectrum and time spectrum of γ-rays can be measured based on DC-TOF method, and the circuit diagram used here is shown in Fig. 5. The multichannel analyzers MCA4 and MCA5 was used to obtain the electron recoil spectra generated by the prompt γ-rays and delayed γ-rays in the detector, and MCA2 and MCA3 were used to get the time spectra of the neutrons and γ-rays.

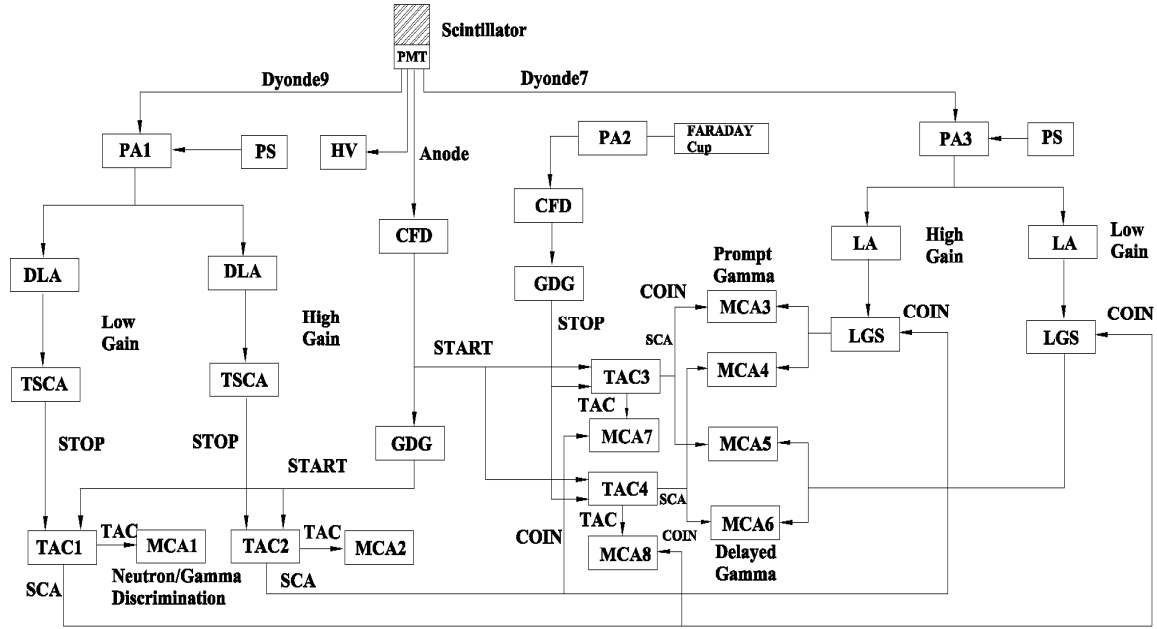

| | | |
|---|---|---|
| PMT | Photomultiplier tube | XP2020 |
| HV | High voltage power supply | 456H |
| PS | Power supply | *ORTEC* | 4002P |
| PA1 | Preamplifier | *CANBERRA* | 2005BT |
| PA2 | Preamplifier | VT120B |
| PA3 | Preamplifier | 113 |
| CFD | Constant fraction discrimination | 584 |
| GDG | Gate & delay generator | 416A |
| DLA | Delay line amplifier | *ORTEC* | 460 |
| TSCA | Timing SCA | 551 |
| TAC | TAC/SCA | 567 |
| LA | Linear Amplifier | 572A |
| LGS | Linear gate and stretcher | 542 |
| MCA | Multichannel analysis | 927 |

Fig. 5. Electronics circuit of DC-TOF method (all the electronic modules are fabricated by *ORTEC* exclude PA1, which is fabricated by *CANBERRA*).

## 4 Results
### 4.1 Time spectra

Three time-to-amplitude-converters (TACs) were adopted in the circuit shown in Fig. 5. TAC1 was used to discriminate the neutrons and γ-rays; TAC2 and TAC3 were used to select the prompt time interval and delayed time interval, respectively. The time calibrator ORTEC 462 (not shown in Fig. 5) was used to calibrate all three TACs. In calibration, the pulse period of time calibrator was set to 80 ns, the dispersion output pulse mode was selected in order to confirm the peak position more accurate, the TAC range was set to 1000 ns, and the multi-channel analyzer was set to 2048 channels. The standard time pulses are shown in Fig. 6, and the time interval per channel can be determined based on the results in Fig. 7.

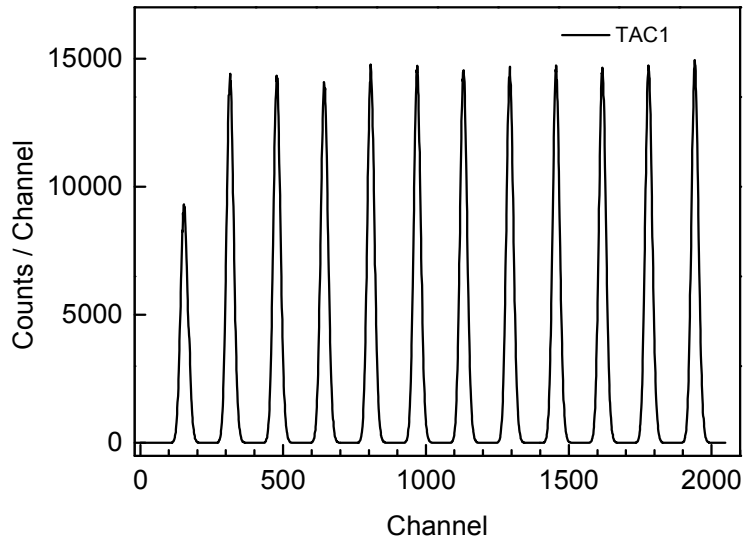

Fig. 6. Standard pulse generated by ORTEC 462.

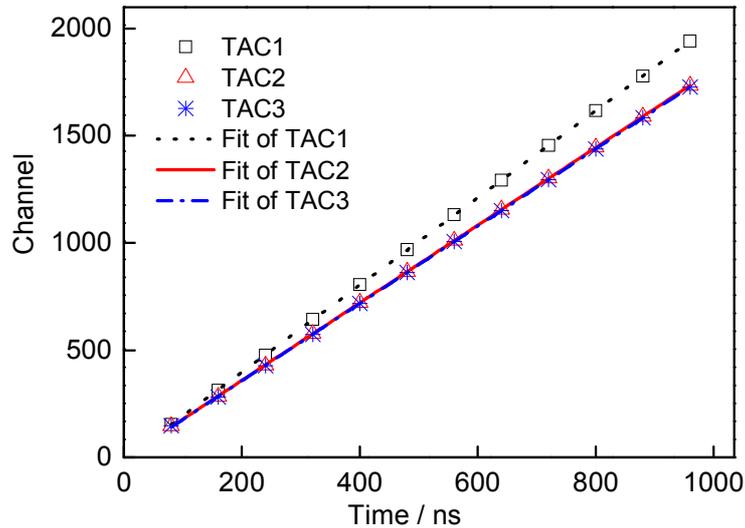

Fig. 7. Linear fitting of the TACs.

The fitting results shown in Fig. 7 give coefficients of (2.0328 ± 0.0018), (1.8094 ± 0.0022) and (1.8022 ± 0.0023) ns/Channel with standard errors of 0.09%, 0.12% and 0.13% corresponding to TAC1 to TAC3, respectively. The correlation coefficient is higher than 0.9999. The deviation of coefficients between TAC2 and TAC3 is only 0.4%.

The duration of prompt γ-ray emission, which is the transportation time of neutrons in the DU shell of 3.1 cm thickness, is very fast. The full width at half-maximum (FWHM) of the neutron pulse is about 10 ns (shown in Fig. 8). Because the measured time spectrum was convolved with the neutron pulse shape, the actual γ-ray time spectrum can be obtained only if the neutron pulse shape can be deconvolved. For that, a deconvolution method was used to eliminate the influence of the pulsed neutron shape. The pulse shape in Fig. 8 was substituted by a Gaussian waveform with FWHM of 10 ns, which has a total probability of 1, generated by the Gaussian probability density function $\rho = \frac{1}{\sqrt{2\pi}\sigma}\exp(-\frac{(E-E_0)^2}{2\sigma^2})$. This Gaussian function was used to generate a Gaussian matrix with a dimension of 70 × 70. Finally, we obtained the prompt γ-ray time spectrum based on the iterative deconvolution method. Relative comparison of experiment to MCNP5 simulation is shown in Fig. 9; they are qualitatively consistent with each other.

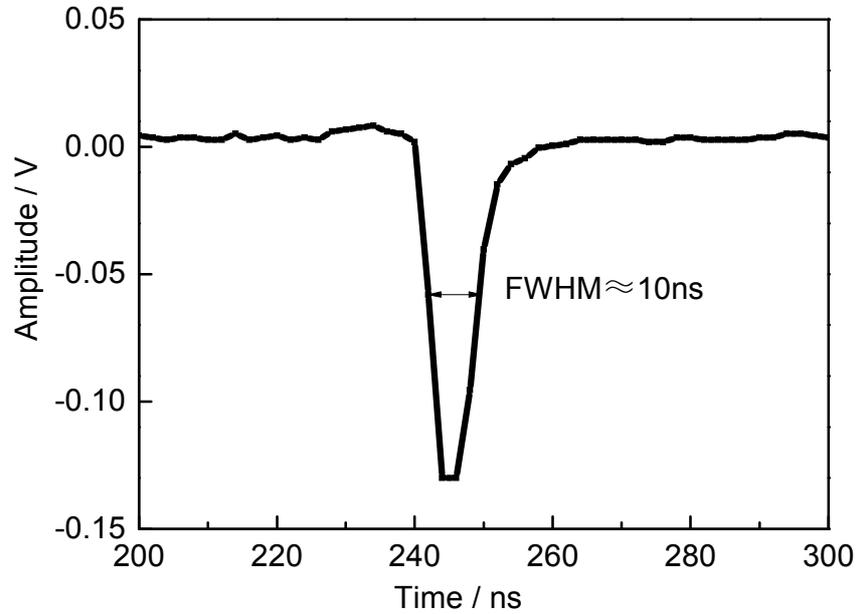

Fig. 8. Pulse shape obtained from Faraday cup.

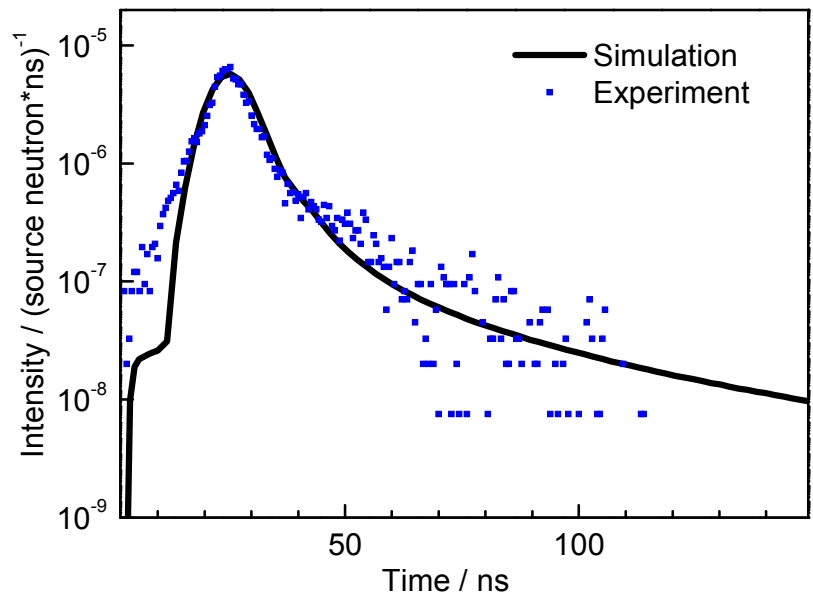

Fig. 9. Time spectra of prompt γ-rays from experiment and MCNP5 simulation.

Because the prompt γ-ray emission process is very short, the assumption of a Gaussian function and the instability of the neutron pulse cause prominent differences between experiment and simulation results. Therefore, their comparison only gives a relative result in this paper. In order to obtain a more accurate γ-ray time spectrum, the actual neutron pulse shape instead of a Gaussian shape should be used in deconvolution. Further increasing the signal intensity and using a better deconvolution method should also be done in future work.

### 4.2 Electron recoil spectra and prompt γ-ray spectra

The γ-rays were measured by a BC501A liquid scintillation detector based on the circuit shown in Fig. 5. The electron recoil spectra of the prompt γ-rays and delayed γ-rays were obtained at the same time. The measuring times of the prompt and delayed γ-rays were 206.5 ns and 600.1 ns, respectively, with a ratio of 2.91. This ratio was also the time correction factor used in subtracting the delayed γ-rays. The electron recoil spectra of prompt, delayed and total γ-rays are shown in Fig. 10. The intensity of prompt γ-rays was 2 to 3 times stronger than that of delayed ones when the energy was below 0.5 MeVee, while their difference was 6 to 30 times when the energy was above 0.5 MeVee.

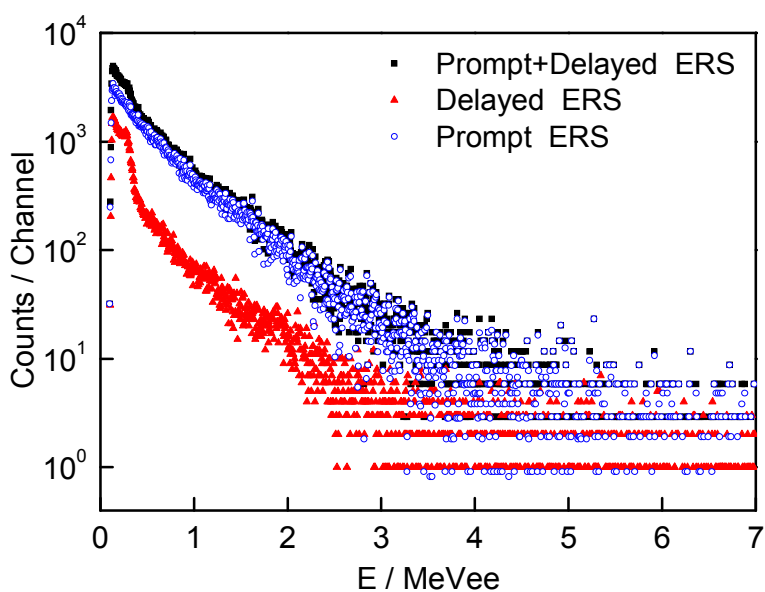

Fig. 10. Prompt, delayed and background subtracted ERS.

The experimental ERS and the simulation based on MCNP5 and EGSnrc Monte Carlo method are shown in Fig. 11. They agree well with each other below 4 MeVee, with the deviations less than 6% and 12% in the energy regions 0.4–2 MeVee and < 3 MeVee, respectively.

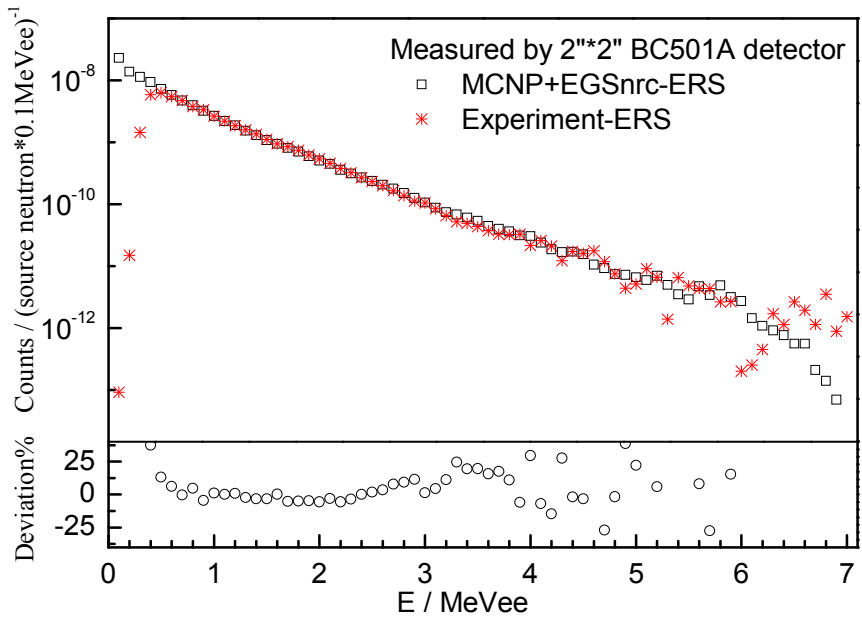

Fig. 11. Comparison between experiment and simulation for ERS.

The iterative method, the Tikho regularization method and inverse matrix method were used to unfold the ERS and generate the prompt γ-ray energy spectra, as shown in Fig. 12 with the simulation result. In the energy region less than 3 MeV, the two agree rather well, with a deviation less than 20% at most points. In the higher energy region, fluctuation of the experimental data is prominent due to the bad statistics and unfolding method, but the trend of the experimental spectrum is consistent with the simulation.

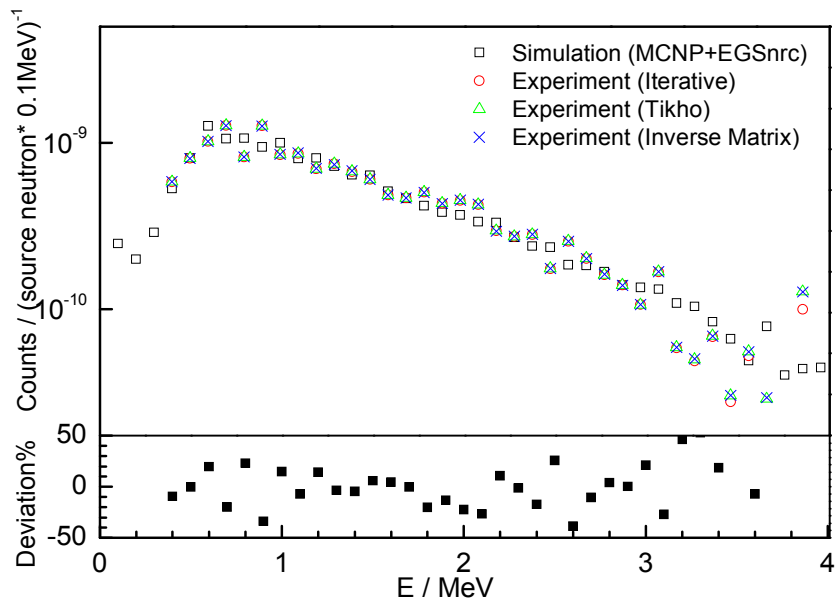

Fig. 12. Comparison between experimental prompt γ-ray spectrum and calculation result.

Because deviations between experiment and simulation are usually significant at some energies, some scholars have used the integral results with various energy regions for comparison

[22, 23]. In the energy region 0.48 to 1.0 MeV and 1.0 to 7.7 MeV, the deviations between Hansen's experiment and simulation with ENDL Library are 5.6% and 13.5%, respectively [22]. The integral results of the present experiment and simulation based on the ENDF/BVI.5 Library are shown in Table 4. In the energy region 0.4 to 1 MeV and 1 to 3 MeV, the deviations are 9.2% and 1.1%, respectively. The result in the 1 to 3 MeV region is better than that of Ref. [22].

Table 4. Integral ratios and deviations of the prompt γ-ray spectra.

| Energy interval | Integral ratio to the whole spectra | Accumulated ratio | Deviation to simulation |
|---|---|---|---|
| MeV | % | % | % |
| <0.4 | 4.4 | 4.4 | — |
| 0.4–1 | 40.3 | 44.7 | 9.2 |
| 1–3 | 48.2 (88.5) | 92.9 | 1.1 (1.3) |
| 3–5 | 6.2 | 99.1 | 30.9 |
| >5 | 0.9 | 100 | — |

As indicated in Table 4, it is very difficult to measure the prompt γ-rays below 0.4 MeV and above 3 MeV, due to the tiny prompt γ-ray intensity. For high-energy prompt γ-rays, the extremely low count rate is also a barrier, while the γ-ray absorption and the limitation of neutron/γ-ray discrimination is a major challenge for low-energy γ-rays. For energies from 0.4 MeV to 3 MeV, which give the contribution of 88.5% to the total signal, rather good experimental results are obtained here.

### 4.3 Uncertainty analysis

In the present work, the uncertainties of simulated ERS and γ-ray time spectrum are estimated to be less than 5%. The uncertainty sources in experiments and the estimated values are shown in Table 5. For the ERS, the uncertainty of the γ-ray response function is determined by the difference between the simulation and the actual response of the standard γ-ray radiation source. Statistical error refers to statistical counts for each energy bin in the ERS. The uncertainty of neutron/γ-ray discrimination mainly depends on the time resolution of the detector, the discrimination method and noise. The uncertainty of the unfolding method refers to the difference between the unfolded spectrum and the ideal spectrum. The uncertainty of neutron source yield is determined by the associated α particles measured by the Au(Si) surface barrier semiconductor detector. The uncertainty of the γ-ray time window is determined by the selection of prompt and delayed time window and the uncertainty in time calibration. The tritium target is in the center of the uranium shell within an error of 1 mm. In view of the distance between the tritium target and the detector being 10.7 m with an uncertainty of less than 1 cm, the uncertainty of distance contributes about 0.5% to the total uncertainty.

For the uncertainty of the γ-ray time spectrum, by comparison with that of the ERS, the neutron pulse shape and time to amplitude conversion items were included, while response function and neutron/γ-ray discrimination were excluded. The uncertainty of time to amplitude conversion is determined by the TAC (ORTEC 567) and time calibrator (ORTEC 462). The uncertainty of the pulse neutron shape refers to the instability, which makes up the main contribution to the total uncertainty.

Table 5. Sources of uncertainty in γ-ray energy spectrum and time spectrum.

| Energy Spectrum | | Time Spectrum | |
| --- | --- | --- | --- |
| Item | Uncertainty% | Item | Uncertainty% |
| Response functions | <5 | Shape of source neutrons | 5−20 |
| Statistic error | <10 | Statistic error | 15 |
| Neutron/γ-ray discrimination | <1 | Time amplitude conversion | <0.5 |
| Background subtracting | 2 | Background subtracting | 2 |
| Unfolding method | <10 | Unfolding method | <10 |
| Neutron source yield | 1.6−2 | Neutron source yield | 1.6−2 |
| γ-ray time window | <1 | γ-ray time window | <1 |
| Source target position | <0.5 | Source target position | <0.5 |

## 5 Conclusions

The method of time of flight and pulse shape coincidence with energy spectrum (DC-TOF) was proposed in this paper. Prompt γ-ray spectra generated by D-T reaction neutron bombarding DU spherical shell were obtained based on the DC-TOF method. The experimental prompt γ-ray time spectrum and the simulation results are fairly consistent. The experimental prompt γ-ray energy spectra are in good agreement with the simulation results based on the ENDF/BVI.5 library in the energy region 0.4−3 MeV. For the comparison of integral quantities, the deviations are 9.2% and 1.1% in the energy regions of 0.4−1 MeV and 1−3 MeV, respectively. The experimental prompt γ-rays have a significant deviation if γ-ray energy is beyond 3 MeV, due to the bad statistics in the higher energy region. The experimental uncertainty is mainly determined by the recoil electron statistical error and the uncertainty of the unfolding method. The measurement technique introduced in this paper can provide a reference for measuring the multiplicities of prompt γ-rays and the $^{238}$U nuclear parameters for test experiments used in hybrid reactors. A digitizer will be adopted to replace the traditional electronics modules in future work, which should improve the stability of data acquisition and the neutron γ-ray discrimination ability.